\documentclass[twocolumn,prb,aps,showpacs,amsmath,amssymb,superscriptaddress]{revtex4}
\usepackage{graphicx}
\usepackage{amssymb}
\usepackage{psfrag}
\usepackage{amsfonts}
\usepackage[T1]{fontenc}
\usepackage[latin1]{inputenc}
\graphicspath{{figures_eps/}}

\usepackage{graphicx}
\usepackage{psfrag}
\usepackage{color}
\usepackage{amsfonts}
\usepackage{amsmath}
\usepackage{amssymb}
\usepackage{ifthen}

\begin{document}

\title{Parity detection and entanglement with a Mach-Zehnder interferometer}

\author{G\'eraldine Haack}
\affiliation{D\'epartement de Physique Th\'eorique, Universit\'e de Gen\`eve, CH-1211 Gen\`eve 4, Switzerland}

\author{Heidi F\"orster}
\affiliation{United Nations University, Hermann-Ehlers-Str. 10, 53113 Bonn, Germany}

\author{Markus B\"uttiker}
\affiliation{D\'epartement de Physique Th\'eorique, Universit\'e de Gen\`eve, CH-1211 Gen\`eve 4, Switzerland}

\date{\today}


\begin{abstract}
A parity meter projects the state of two qubits onto two subspaces with different parities, the states in each parity class being indistinguishable. It has application in quantum information for its entanglement properties. In our work we consider the electronic Mach-Zehnder interferometer (MZI) coupled capacitively to two double quantum dots (DQDs), one on each arm of the MZI. These charge qubits couple linearly to the charge in the arms of the MZI. A key advantage of an MZI is that the qubits are well separated in distance so that mutual interaction between them is avoided. Assuming equal coupling between both DQDs and the arms and the same bias for each DQD, this setup usually detects three different currents, one for the odd states and two for each even state. Controlling the magnetic flux of the MZI, we can operate the MZI as a parity meter: only two currents are measured at the output, one for each parity class. In this configuration, the MZI acts as an ideal detector, its Heisenberg efficiency being maximal. Initially unentangled DQDs become entangled through the parity measurement process with probability one and for a class of initial states our parity meter deterministically generates Bell states. 
\end{abstract}

\maketitle

\section{Introduction} 
A parity meter couples two qubits and allows to measure whether the qubits are in the even subspace spanned by the even states $\big\{\vert {\uparrow} {\uparrow} \rangle, \vert{\downarrow} {\downarrow} \rangle\big\}$ or in the odd subspace spanned by the odd states $\big\{\vert {\uparrow} {\downarrow} \rangle, \vert{\downarrow} {\uparrow} \rangle\big\}$. A particular basis for these two subspaces are the four Bell states, which are the maximal entangled states for two qubits. As these Bell states are also eigenstates of the parity operator for two qubits, a parity measurement drives the qubits into a Bell state and constitutes a source of entanglement. The parity detection is thus of special importance in the possible implementation of quantum computation schemes.\cite{Beenakker04, Engel05,Trauzettel06, Zilberberg08}

\begin{figure}[ht!]
\centering
\includegraphics[width=6cm]{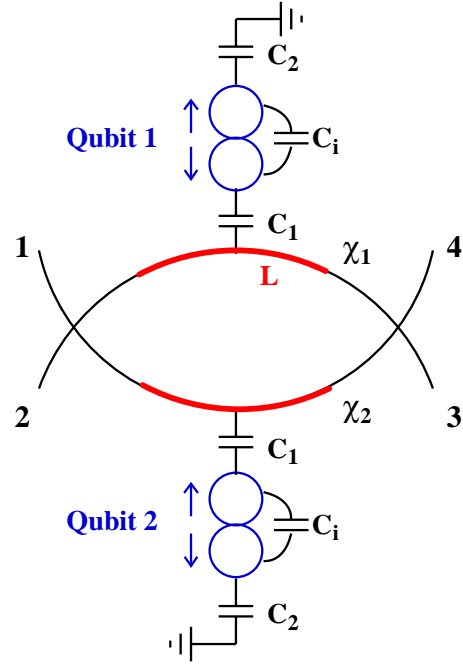} \caption{Mach-Zehnder interferometer (MZI) coupled capacitively to two qubits. $\chi_1 (\chi_2)$ is the phase accumulated by the electrons passing through the upper (lower) arm of the MZI over a length L. These phases depend on the state of the qubits. $C_1, C_2$ and $C_\text{i}$ are the capacitances characterizing the double quantum dots (DQDs).}
\label{fig1}
\end{figure}

Realizations of parity meters for charge and spin qubits have been discussed considering a quantum point contact (QPC) as detector.\cite{Trauzettel06, Loss06} The purpose of our work is to investigate parity detection of charge qubits with the help of a Mach-Zehnder interferometer (MZI) as shown in Fig.\ \ref{fig1}. The MZI is the simplest possible type of interferometer. It has recently been implemented in a two-dimensional electron gas in the quantum Hall regime with electron motion along edge states.\cite{Heiblum03, Neder06, Roulleau08, Litvin08, Bieri09, Roulleau09} A discussion which compares the MZI with a second order two-particle interferometer is presented in Ref. \onlinecite{Chung05}. The non-linear behavior of the MZI has received much 
experimental \cite{Neder06, Roulleau08, Litvin08, Bieri09} and theoretical attention.\cite{Sukhorukov07, Levkivskyi08, Chalker09_1, Chalker09} In the following we assume that the applied voltages are small and the MZI is operated within its linear response regime.\cite{note2} The MZI coupled to qubits in its arms has the advantage that the two qubits are (on the mesoscopic scale) far away from each other. Therefore they are not subject to mutual Coulomb interaction. An additional advantage is that the MZI is tunable both through a variation of the applied magnetic field and through gate voltages. These gate voltages allow to determine the length of the interferometer arms and to control the transparency of the QPCs which act as beam splitters. Admittedly the fabrication of such a sample is not simple since one of the qubits would have to be built in the interior of the Corbino disk. As a consequence its gates can only be biased with wires which will bridge the conducting region of the interferometer. 

Our analysis of the parity detection is based on a formulation which takes into account that a quantum measurement lasts a finite time. This assumes that the detector is only weakly coupled to the system. Such weakly coupled detectors are typical for on chip detectors in mesoscopic physics and have led to a strong theoretical development.\cite{Levinson97,Gurvitz97, Korotkov99,Buttiker00,Korotkov01, Milburn01, Korotkov02,Pilgram02, Clerk03, Averin05, Kang06, Korotkov06, Jordan06, Nigg09, Clerk09, Clerk10} 
Previous works have shown that linearly or quadratically weakly coupled detectors such as QPCs distinguish in general three currents. Their measurement process allow purification of a mixed state and entanglement of the qubits.\cite{Ruskov03, Mao04, Ruskov06, Jordan08} In subsequent work Trauzettel \textit{et al.} \cite{Trauzettel06} have shown how to use a QPC as a parity detector. The investigation of entanglement generation finds that  the detector still distinguishes three states \cite{Trauzettel06}, two states which are detectable in repeated measurements belonging to the two parity classes and a time-dependent state which oscillates between the parity classes. These results follow from a qubit Hamiltonian formulated in the $\hat{\sigma}_x $-basis. In this work we treat qubits without tunneling and show that the detector can be set to measure only two currents. 

The paper is organized as follows: In Section II, we describe our setup (see Fig.\  \ref{fig1}).  In Section III, we give the conditions for the MZI to operate as parity meter and in Section IV, we investigate its Heisenberg efficiency. The efficiency is determined by the ratio of the measurement time of the detector and the dephasing time of the qubits.\cite{Pilgram02, Clerk03, Kang06} We find that a properly tuned MZI parity detector has maximal efficiency, i.e. it is a quantum limited detector. In Section V, we show that the measurement performed by the MZI as parity meter entangles the DQDs with probability one when each DQD is initially in a superposition state. We also investigate in this last section another interesting property of this parity detection: for this particular class of initial states, if the odd current is measured, we know into which odd Bell state the qubits are driven. This makes the MZI an efficient parity detector and, for initial superposed states, a deterministic source of entanglement.

\section{Model} The electronic MZI is built in a quantum Hall bar by using two QPCs as beam splitters. The qubits are spatially separated so that they do not interact directly with each other. The interaction takes place between each qubit and the charges of the corresponding arm and is described with the help of geometrical capacitance coefficients, see Fig.\  \ref{fig1}: $C_1$ determines the coupling of the charge on the DQD to the charge on the arm, $C_i$ determines the coupling between the dots, and $C_2$ is the capacitance of the DQD to the ground. For simplicity, here, the capacitances are assumed to be the same for each qubit. The regime of weak coupling between the detector and the DQDs is defined by $C_1 \ll C_2, C_\text{i}$.  Additionally, to control the system, we can tune the Aharonov-Bohm flux, the length of the arms, and the transmission and reflection probabilities of the QPCs. The Hamiltonian of this model reads
\begin{equation} \label{hamiltonian}
\hat{H}= \hat{H}_{\text{qb}} + \hat{H}_{\text{int}} + \hat{H}_{\text{det}},
\end{equation}
where
\begin{eqnarray}
\hat{H}_{\text{qb}} &=&  \frac{\epsilon_1}{2} \hat{\sigma}_z^1 + \frac{\epsilon_2}{2} \hat{\sigma}_z^2 + \frac{\Delta_1}{2}  \hat{\sigma}_x^1 + \frac{\Delta_2}{2}  \hat{\sigma}_x^2  \label{H_{qb}}, \\
\hat{H}_{\text{int}} &=& \hat{U}_1(\frac{1-\hat{\sigma}_z^1}{2}) + \hat{U}_2(\frac{1+\hat{\sigma}_z^2}{2}) \label{H_{int}}.
\end{eqnarray}
$\hat{H}_{\text{det}}$ is the Hamiltonian of the MZI. $\hat{H}_{\text{qb}}$ is the Hamiltonian of the qubits. We assume no tunneling between the dots, $\Delta_1 = \Delta_2 = 0$, and the same bias for the two DQDs, $\epsilon_1 = \epsilon_2 = \epsilon$, so that $\hat{H}_{\text{qb}} = (\epsilon/2) (\hat{\sigma}_z^1 + \hat{\sigma}_z^2)$. The Pauli matrices for the upper and lower qubits are $\hat{\sigma}_z^1$ and $\hat{\sigma}_z^2$. In the interaction Hamiltonian $\hat{H}_{\text{int}}$ the  expressions $(1-\hat{\sigma}_z^1)/2$ and $(1+\hat{\sigma}_z^2)/2$ are the charge operators of the qubits which couple to the charge on the arms of the interferometer. The Mach-Zehnder detector considered here is a linear detector in the sense that qubits couple linearly to the charge in the arms of the MZI through the $\hat{\sigma}_z$ operator. As the qubit Hamiltonian is also in the $\hat{\sigma}_z $-basis, there is no relaxation process for the DQDs due to the coupling with the detector. As shown in more detail in the next section investigating the MZI as parity meter, the parity operator $\hat{P} = \hat{\sigma}_z^1 \otimes \hat{\sigma}_z^2$ commutes with both $\hat{H}_{\text{qb}}$ and $\hat{H}_{\text{int}}$. It follows that, in this configuration, the Mach-Zehnder detector is a quantum non demolition detector (QND-detector) of the parity classes.\cite{Caves80, Braginsky92, Haroche06} This is to say that repeated measurements will find the qubits in the same parity class, the MZI detects only two stationary currents. This is an important distinction between previous works on entanglement generation which use a qubit Hamiltonian in the $\hat{\sigma}_x$-basis \cite{Ruskov03, Mao04, Ruskov06, Trauzettel06} and the  investigation presented here.  To precise $\hat{H}_{\text{int}}$, $\hat{U}_{1,2}$ are the potential operators on the upper and lower arms respectively. They depend on the bare charge and on the screening charge on the arms due to the presence of the DQDs.\cite{Buttiker00, ButtikerKu} This interaction Hamiltonian implies that the electrons transiting in the arms of the MZI are sensitive to the charge in the DQDs only if this charge is in the dot close to the arm. As a consequence, electrons on arm $i=1,2$ will acquire during their transfer a phase $\chi_i$ due to the magnetic flux, due to the length of the arm $i$ and due to the state of the DQDs:
\begin{eqnarray}
\chi_1 &=& k(d-L) + \theta_{up} + \Delta \chi_1 \frac{1-\hat{\sigma}_z^1}{2},  \label{phase1}\\
\chi_2 &=&  k(d-L) + \theta_{d} + \Delta \chi_2 \frac{1+\hat{\sigma}_z^2}{2} \label{phase2}.
\end{eqnarray}
$\Delta \chi_i$ is the additional phase depending on the state of the DQDs and $\theta_{up} - \theta_d = 2\pi \Phi / \Phi_0$ is the magnetic flux between the arms. For simplicity we consider both arms having the same length $d$ and the electrons are sensitive to the DQDs over a length $L$, $k$ is the Fermi wavevector, see Fig.\ \ref{fig1}. Equations\ (\ref{phase1}) and (\ref{phase2}) should be viewed only as abbreviations, since in reality the phases and the transmission probability $T_{31}$ to be given in the next section are c-numbers depending on the state of the qubits. Thus instead of writing operators in Eqs.\ (\ref{phase1}) and (\ref{phase2})  we should index these quantities with the corresponding qubit states. Here we avoid such a heavy notation and keep operators. 

In the next sections, we will derive the conditions under which the MZI coupled to DQDs allows their parity detection and operates as ideal detector in this regime. We will then investigate the creation of entanglement through the parity measurement.

\section{Parity meter} 

A parity meter is a quantum detector which can only distinguish states belonging to different parity classes, the odd and even classes. The observable we measure is the averaged current in lead 3 (see Fig.\ \ref{fig1}). Assuming that the electrons are injected in lead 1, the current is given by \cite{Chung05}
\begin{equation}
I = \frac{2e^2V}{h}T_{31}.
\end{equation}
$T_{31}$ is the transmission probability from lead 1 to lead 3, 
\begin{equation} \label{transmission}
T_{31} = \big(T_L R_R + T_R R_L + 2\sqrt{T_R R_LT_L R_R} \cos \tilde{\Phi} \big).
\end{equation}
It can be calculated with the help of the scattering matrix $S$ of this model: $T_{31}=\vert S_{31}\vert^2$, where $S_{31}$ is one element of the $S$-matrix. It corresponds to the amplitude probability of going from lead 1 to lead 3,
\begin{equation} \label{S_31}
S_{31} = \sqrt{T_L} e^{i\chi_2} \sqrt{R_R} + \sqrt{R_L} e^{i\chi_1}\sqrt{T_R}.
\end{equation}
We introduce the phase $\tilde{\Phi} = \chi_1 - \chi_2$ where $\chi_1$ and $\chi_2$ are given in Eqs.\ (\ref{phase1}) and (\ref{phase2}). Depending on the four possible states of the qubits $\{\vert {\uparrow} {\uparrow} \rangle, \vert {\uparrow} {\downarrow} \rangle, \vert {\downarrow} {\uparrow} \rangle, \vert {\downarrow} {\downarrow} \rangle \}$, $\tilde{\Phi}$ can take four different values: 
\begin{eqnarray}
\tilde{\Phi}_{{\uparrow} {\uparrow}} &=& 2\pi \Phi/\Phi_0 -\Delta \chi_2,  \label{current_phase_1}\\
\tilde{\Phi}_{{\uparrow} {\downarrow}} &=& 2\pi \Phi/\Phi_0, \label{current_phase_2} \\
\tilde{\Phi}_{{\downarrow} {\uparrow}} &=& 2\pi \Phi/\Phi_0 + \Delta \chi_1 -\Delta \chi_2, \label{current_phase_3} \\
\tilde{\Phi}_{{\downarrow} {\downarrow}} &=& 2\pi \Phi/\Phi_0 + \Delta \chi_1.  \label{current_phase_4}
\end{eqnarray}
To operate the MZI as parity meter, only two currents have to be distinguishable: $I_o$ corresponding to the odd states $\big\{\vert {\uparrow} {\downarrow} \rangle, \vert{\downarrow} {\uparrow} \rangle\big\}$ and $I_e$ corresponding to the even states $\big\{\vert {\uparrow} {\uparrow} \rangle, \vert{\downarrow} {\downarrow} \rangle \big\}$. Considering Eqs.\ (\ref{current_phase_2}-\ref{current_phase_3}), only one current $I_o$ for the odd states will be measurable if the incremental phases $\Delta \chi_i$ for electrons in upper and lower arms are the same:
\begin{equation} \label{parity1}
\Delta \chi_1 = \Delta \chi_2 = \Delta \chi.
\end{equation}

Assuming Eq.\ (\ref{parity1}) and measuring the output current as a function of the magnetic flux, three currents will be in general distinguishable: one for the odd states $I_o$ and two for the even states $I_{{\uparrow}{\uparrow}}$ and $I_{{\downarrow}{\downarrow}}$ as shown by the three black points on Fig.\ \ref{current}\,a. Although this measurement is not a true parity measurement, it already allows entanglement between the qubits as investigated with a QPC coupled to two DQDs.\cite{Ruskov03, Mao04, Trauzettel06} In our work we want a true parity meter with advantages for the entanglement probabilities as discussed in the next section. Equations\ (\ref{current_phase_1}) and (\ref{current_phase_4}) and Fig.\ \ref{current}\,b show that measuring only one current $I_e$ for the even states is possible if the magnetic flux is set to 
\begin{equation} \label{parity2}
2\pi \Phi/\Phi_0 = 0 \quad \text{modulo}\quad \pi,
\end{equation}
providing a maximum or minimal transmission. To be specific, in the reminder of the work, we will choose $2\pi \Phi/\Phi_0 = 2\pi$ to be at the maximum of transmission.

\begin{figure}[h!]
\centering
\includegraphics[width=6.5cm]{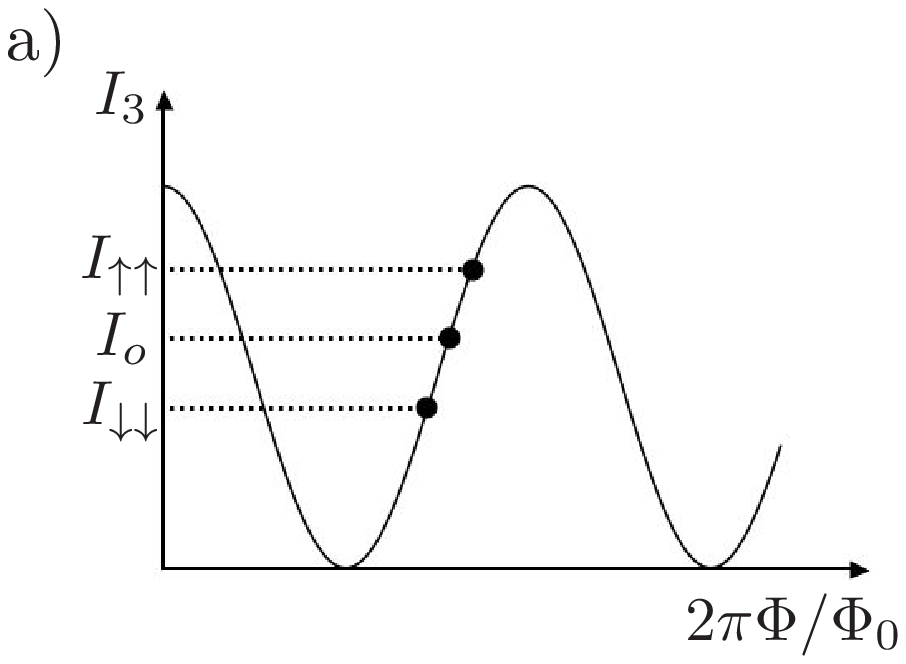}
\end{figure}
\begin{figure}[h!]
\centering
\includegraphics[width=6.5cm]{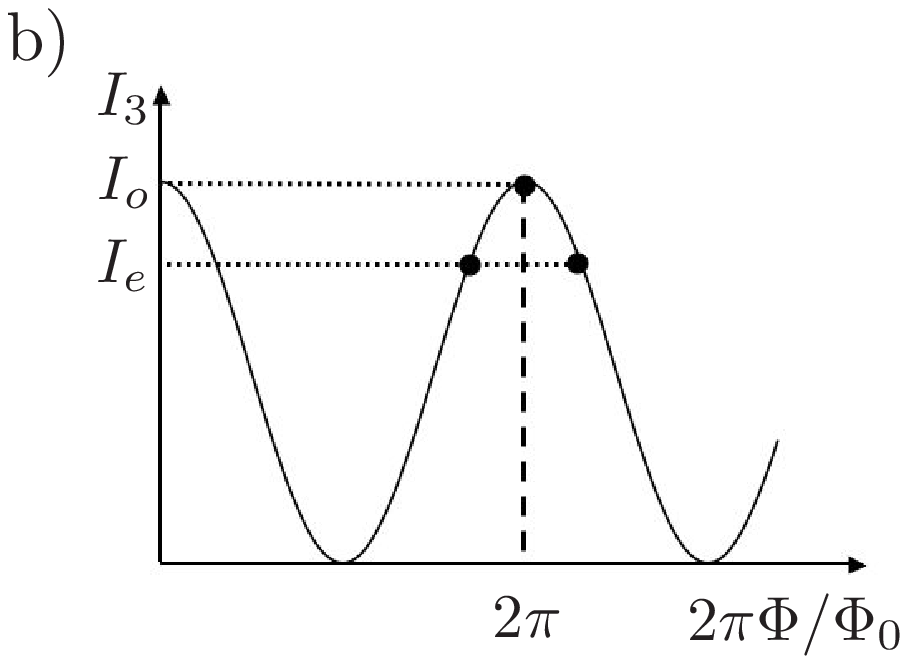}
\caption{Current measured at lead 3 as a function of the magnetic flux between the arms of the MZI. a) At an arbitrary flux three currents $I_o,I_{{\uparrow}{\uparrow}}$ and  $I_{{\downarrow}{\downarrow}}$ are distinguishable even if the incremental phases are identical: $\Delta \chi_1\! = \!\Delta \chi_2\! =\! \Delta \chi.$ b) To measure only two currents corresponding to the parity classes, $I_o$ and $I_e$, we set the flux to  $2\pi \Phi/\Phi_0 \!= \! 0 \!\quad \! \text{modulo}\! \quad\! \pi$. A maximum of the current is reached at $2\pi \Phi/\Phi_0 = 2\pi$.}
\label{current}
\end{figure}

Equations\ (\ref{parity1}) and (\ref{parity2}) are the two conditions required for the MZI to act as parity meter. Then the transmission probability $T_{31}$ can be expanded around $2\pi \Phi/\Phi_0 = 2 \pi$ up to second order in $\Delta \chi$. It depends then on the states of the qubits {\it only} through the parity operator $\hat P = \hat{\sigma}_z^1\otimes \hat{\sigma}_z^2$:
\begin{eqnarray}
\!T_{31} \!&=& \!T_L R_R + T_R R_L  \nonumber \\
\!&&\! + 2 \sqrt{T_R R_LT_L R_R}\! \Big( 1-\frac{\Delta \chi^2}{4}(1\!+\!\hat{\sigma}_z^1 \!\otimes \!\hat{\sigma}_z^2)\!\Big).
\end{eqnarray}

This measurement operator $\hat{P}$ allows parity measurement and the creation of entanglement. Indeed $\hat{P}$ has two eigenvalues $\pm 1$ of degeneracy 2, corresponding respectively to the even parity class and to the odd parity class. Its four eigenstates are the Bell states,
\begin{eqnarray}
\vert \psi_e^{+} \rangle \!&=&\! \frac{1}{\sqrt{2}}\left(\!\vert {\uparrow} {\uparrow} \rangle \!+\! \vert {\downarrow} {\downarrow} \rangle \!\right)\!, \quad
\vert \psi_e^{-} \rangle \!=\! \frac{1}{\sqrt{2}}\left(\!\vert {\uparrow} {\uparrow} \rangle \!-\! \vert {\downarrow} {\downarrow} \rangle \!\right)\!, \! \label{Bell_even} \\
\vert \psi_o^{+} \rangle \!&=&\! \frac{1}{\sqrt{2}}\left(\!\vert {\uparrow} {\downarrow} \rangle \!+\! \vert {\downarrow} {\uparrow} \rangle \!\right)\!, \quad
\vert \psi_o^{-} \rangle \!=\! \frac{1}{\sqrt{2}}\left(\!\vert {\uparrow} {\downarrow} \rangle \!-\! \vert {\downarrow} {\uparrow} \rangle \!\right)\!. \label{Bell_odd}
\end{eqnarray}
The lower indices indicate the parity class and the upper indices ($\pm$) indicate whether the sum or difference of the computational states is taken. 
These Bell states are of particular interest as they share the properties to be the maximal entangled states for two qubits and to have a definite parity. A key property of the qubit Hamiltonian $\hat{H}_{\text{qb}}$ considered in this work is that its dynamics conserves the parity of the Bell states:
\begin{equation} \label{qubit_eigen}
\hat{H}_{\text{qb}} \vert \psi_e^{\pm} \rangle = \epsilon \vert \psi_e^{\mp} \rangle \quad \text{and} \quad
\hat{H}_{\text{qb}} \vert \psi_o^{\pm} \rangle = \epsilon \vert \psi_o^{\pm} \rangle.
\end{equation}
It is the parity measurement in the $\hat{\sigma}_z$-basis combined with the qubit Hamiltonian $\hat{H}_{\text{qb}}$ and the coupling in this $\hat{\sigma}_z$-basis (no relaxation process) that allows the MZI to act as parity meter for the DQDs during the entire measurement process. In this sense, this weak measurement procedure is the analog of a QND measurement of parity classes.\\

In the next section we will investigate the efficiency of the MZI as parity meter to show that it acts as an ideal detector. The efficiency $\eta$ is defined as the ratio of the dephasing time over the measurement time, $0<\eta \leq 1$. If $\eta =1$, the quantum detector is an ideal detector: it extracts information on the quantum system as fast as it dephases it. In other words the measurement process is the only dephasing source. The ideality of the detector will constitute a key property for the success of entanglement generation (Section V).

\section{Efficiency} 
The quantum efficiency $\eta$ is defined as $\eta = \Gamma_{\text{m}} / \Gamma_{\text{deph}}$ , where $\Gamma_{\text{deph}}$ is the dephasing (or decoherence) rate of the measured quantum system and $\Gamma_{\text{m}}$ is the measurement rate of the detector. The most efficient quantum detector one can implement is a quantum-limited detector, $\eta=1$; it acquires the information about the measured system as fast as this quantum system loses its quantum coherence. $\Gamma_{\text{deph}}$ is the inverse of the coherence time $T_2$ of the measured quantum system and $\Gamma_{\text{m}}$ is defined as the inverse of the time needed by the detector to distinguish the signal from the output noise,\cite{Korotkov99}
\begin{equation} \label{meas_rate}
\Gamma_{\text{m}} = \frac{(\Delta I)^2}{4S_{II}}.
\end{equation}
$S_{II}$ is the low-frequency noise spectrum \cite{Buttiker92} and $\Delta I = I_o - I_e$ is the difference between the even and odd averaged currents measured in lead 3 (see Fig.\ \ref{current}\,b). Assuming linear response, $\Delta I \ll I_o, I_e$, the shot noise is independent of the qubits' states:
\begin{equation} \label{noise}
S_{II} = \frac{S_e + S_o}{2}, \quad S_{e,o} = \frac{2 e^3 V}{h} T_{31,e,o} (1 - T_{31,e,o}).
\end{equation}
$T_{31,e,o}$ are the transmission probabilities for the even and odd configurations respectively. They are given by Eq.\ (\ref{transmission}) in which the parity operator $\hat{P}$ takes the values $\pm1$ depending on the parity class. The current difference up to the second order in the small parameter $\Delta \chi$ is
\begin{equation}
\Delta I = \frac{2 e^2 V}{h} \sqrt{R_RT_RR_LT_L} (\Delta \chi) ^2.
\end{equation}
The maximal value of $\Gamma_{\text{m}}$ is reached for $R_L = R_R = 1/2$:
\begin{equation}
\Gamma_{\text{m}}(\!R_L\! =\! R_R\! =\! 1/2\!) = \frac{1}{8} \frac{eV}{h} \Delta \chi^2 = \frac{1}{8} \frac{eV}{h} \left(\frac{C_\mu}{C_\text{i}}\right)^2.
\end{equation}
Here the additional phase $\Delta \chi$ is expressed as a function of the capacitances of the system, $\Delta \chi = C_\mu / C_\text{i}$ where $C_\mu^{-1} = C_\text{i}^{-1} + C_1^{-1} + C_2^{-1} + C_q^{-1}$ is the electrochemical capacitance of each DQD. $C_q$ is the quantum capacitance, it depends on the density of states in the arms of the MZI.\cite{Pilgram02,Pilgram04} This will allow us to compare this measurement rate to the dephasing rate of the qubits.\\
In general, for two qubits, there are six dephasing rates describing the coherence between the states $\{\vert {\uparrow} {\uparrow} \rangle, \vert {\uparrow} {\downarrow} \rangle, \vert {\downarrow} {\uparrow} \rangle, \vert {\downarrow} {\downarrow} \rangle \}$.  One way to calculate them is to extract their expressions  from the off-diagonal terms of the density matrix.\cite{Hackenbroich00, Clerk03,Averin05,Kang06} Another way consists in studying the phase fluctuations of the DQDs arising from the fluctuating potentials in the arms.\cite{Buttiker00, Pilgram02,Clerk09} Both approaches lead to the same rates and the details of the two calculations are given in the appendix, Sections A and B. Taking into account the parity assumptions, Eqs. (\ref{parity1}) and (\ref{parity2}), the six rates reduce to three rates between the parity classes:
\begin{eqnarray}
\Gamma_{eo} &=& \frac{1}{2} R_L T_L \frac{eV}{h}\left(\frac{C_\mu}{C_\text{i}}\right)^2, \label{rates1}\\
\Gamma_{ee} &=& 2 R_L T_L \frac{eV}{h}\left(\frac{C_\mu}{C_\text{i}}\right)^2, \\
\Gamma_{oo} &=&  0. \label{rates2}
\end{eqnarray}
$\Gamma_{ij}$ is the dephasing rate between the parity classes $i$ and $j$, where $i$ and $j$  are either $e$ for the even class or $o$ for the odd class. These results reflect the perfect anticorrelation between the events on the upper and lower arms of the MZI due to current conservation in the system,\cite{Seelig01} see also Appendix B.\\
As we are investigating the MZI as parity meter, the only dephasing rate of interest for the efficiency of the detector is the dephasing rate between the even and odd classes $\Gamma_{eo}$. For simplicity we do not give the general expression of the efficiency but we only specify it for the particular case of $R_L = R_R = 1/2$ for which the measurement rate is maximal. For half-transmitting QPCs, the efficiency becomes maximal:
\begin{equation}
\eta_{\text{max}} = \frac{\Gamma_{\text{m}}(\!R_L\! =\! R_R\! =\! 1/2\!)}{\Gamma_{eo}(\!R_L\! =\! 1/2\!)} =1.
\end{equation}

The MZI coupled to two DQDs can thus operate at the quantum limit if both QPCs are half-transmitting, $R_L = R_R = 1/2$. As shown in the next section, the ideality of the MZI will reduce the possible sources of decoherence during the measurement process, enhancing thus the probabilities to get entangled states. These probabilities will be extracted from the density operator, calculated by solving a stochastic differential equation for the evolution of the qubits.

\section{Creation of entanglement} 

The entanglement of the DQDs through the measurement process is investigated by solving the differential stochastic equation governing the evolution of the DQDs:\cite{Korotkov99, Korotkov02, Jordan06}

\begin{eqnarray} \label{SDE}
\dot{\rho}_{ij} &=& - \frac{i}{\hbar} [\hat{H}_{\text{qb}}, \rho]_{ij} - \Big( \frac{(I_i - I_j)^2}{4S_{II}} + \gamma_{ij} \Big) \rho_{ij} \nonumber \\
&& + \xi(t) \Big( I_i + I_j -2 \sum_{k=1}^{4}\rho_{kk}I_k \Big) \frac{\rho_{ij}}{S_{II}}. 
\end{eqnarray}
The current in lead 3 is given by $I(t) = \sum_{k} \rho_{kk}I_k + \xi(t)$, where $\rho_{kk}$ is the density matrix's element associated with the state $\vert \psi_k \rangle$. The indices $k=1,2,3,4$ correspond to the four Bell states  $\{\vert \psi_e^{+} \rangle, \vert \psi_e^{-} \rangle, \vert \psi_o^{+} \rangle, \vert \psi_o^{-} \rangle\}$ defined in Eqs.\ (\ref{Bell_even}) and (\ref{Bell_odd}). Working with the Bell states presents a strong advantage when investigating parity measurements and entanglement as shown in the previous section. $I_i$ corresponds to the average current the state $\vert \psi_i \rangle$ would produce. As the MZI acts as parity meter, $I_1 = I_2 = I_e$ and $I_3 = I_4 = I _o$, where $I_e$ and $I_o$ are the even and odd currents. $\xi(t)$ is the random white shot noise of the output characterized by $\langle \xi(t) \xi(0) \rangle = 2 S_{II} \delta(t)$, where $S_{II}$ is the low-frequency noise spectrum of the MZI (see Eq.\ (\ref{noise})). %$\gamma_{ij}$ is the dephasing rate which depends on the efficiency $\eta$ of the detector.
\\
The differential equation Eq.\ (\ref{SDE}) can be decomposed into three terms. The first term describes the standard Schr\"{o}dinger evolution due to the qubit Hamiltonian $\hat{H}_{\text{qb}}$. The possible sources of decoherence for the qubits are considered in the second term of this equation. They consist of the term $(I_i - I_j)^2/(4S_{II})$ (compared Eq.\ (\ref{meas_rate})) and a term $\gamma_{ij}$ which is finite only if the detector is not ideal. As shown in the previous section, the electronic MZI as parity meter is an ideal detector so $\gamma_{ij}=0$. The only source of decoherence is thus the measurement process. The last term describes the backaction of the measurement result on the qubits evolution through the noise $\xi(t)$ of the current. In this work, we are not interested in the entanglement genesis \cite{Jordan08} so we look at the time-averaged corresponding equation:
\begin{equation} \label{master_eq}
\dot{\rho}_{ij} = - \frac{i}{\hbar} [\hat{H}_{\text{qb}}, \rho]_{ij} -  \frac{(I_i - I_j)^2}{4S_{II}} \rho_{ij}.
\end{equation}
This equation can be solved analytically. In our setup, considering a qubit Hamiltonian without tunneling,  a classical initial state (a fully mixed state) will remain classical. Therefore we investigate an initial state in which the individual qubits are in superposition states, 
 \begin{equation} \label{DQDs_{general}}
\vert \Psi_{qb} \rangle =  \left(\alpha \vert {\uparrow}\rangle + \beta \vert {\downarrow}\rangle \right) \otimes \left( \gamma \vert {\uparrow}\rangle + \delta \vert {\downarrow}\rangle \right),
\end{equation}
where $\alpha, \beta, \gamma, \delta$ are complex numbers such that $\vert \alpha \vert^2 + \vert \beta \vert^2 =  \vert \gamma \vert^2 + \vert \delta \vert^2 =1$. By definition, the diagonal elements of the density operator $\rho_{ii}(t)$ determine the probabilities to be in the state $\vert \psi_i \rangle$ at time $t$. Thus the probability $P_i(t)$ to be in the Bell state $\vert \psi_i \rangle$ will be given by $\rho_{ii}(t)$, the probability to be in the even subspace is given by $P_e(t) = \rho_{11}(t) + \rho_{22}(t)$ and the probability to be in the odd subspace is given by $P_o(t) = \rho_{33}(t) + \rho_{44}(t)$. These probabilities depend in general on the initial state of the two qubits given by Eq.\ (\ref{DQDs_{general}}). From the analytical solutions of Eq.\ (\ref{master_eq}), we find the probabilities to measure outcomes in the different parity subspaces:
\begin{eqnarray} 
P_e &=& \vert \alpha \vert^2\vert \gamma \vert^2 + \vert \beta \vert^2\vert \delta \vert^2, \label{P_{even}}\\
P_o &=& \vert \alpha \vert^2\vert \delta \vert^2 + \vert \beta \vert^2\vert \gamma \vert^2. \label{P_{odd}}
\end{eqnarray}
As expected from the qubit Hamiltonian that does not mix the parity classes, Eqs.\ (\ref{P_{even}}) and (\ref{P_{odd}}) do not depend on time. The electronic MZI in this setup works as perfect parity meter during the entire measurement process. Each time an electron transits through the arms, the DQDs are weakly projected onto a subspace with a definite parity. In our QND-limit (see Section III), we can approximate the weak measurement with many projective measurements as the backaction does not take place.

The probabilities for the qubits to be in one of the Bell states are given by:
\begin{eqnarray}
P_e^+(t) \!&=&\! \frac{\vert \alpha \gamma \!+\! \beta \delta \vert^2}{2} \!\cos^2\Big(\frac{\epsilon t}{\hbar}\Big) \!+\! \frac{\vert \alpha \gamma \!-\! \beta \delta \vert^2}{2} \!\sin^2\Big(\frac{\epsilon t}{\hbar}\Big), \\
P_e^-(t) \!&=&\! \frac{\vert \alpha \gamma\!+\! \beta \delta \vert^2}{2}\! \sin^2\Big(\frac{\epsilon t}{\hbar}\Big) \!+\! \frac{\vert \alpha \gamma \!-\! \beta \delta \vert^2}{2}\! \cos^2\Big(\frac{\epsilon t}{\hbar}\Big), \\
P_o^+ &=& \frac{\vert \alpha \delta \!+\! \beta \gamma \vert^2}{2}, \\
P_o^- &=& \frac{\vert \alpha \delta \!-\! \beta \gamma \vert^2}{2}.
\end{eqnarray}
In the odd subspace the probabilities to entangle the qubits depend only on the initial state via the complex coefficients $\alpha, \beta, \gamma$ and $\delta$. This result is again expected from the qubits' dynamics which induces no evolution in the odd subspace, the odd Bell states being eigenstates of $\hat{H}_{\text{qb}}$ (see Eq.\ (\ref{qubit_eigen})). As a consequence, by choosing specific complex coefficients for the initial state, one can know with probability one in which odd Bell state the qubits are if the odd current is measured. The evolution is more fussy in the even subspace. Although the qubit Hamiltonian conserves parity, it induces an evolution between the two even Bell states due to the bias $\epsilon$:
\begin{eqnarray}
\hat{H}_{\text{qb}} \vert \psi_e^{+} \rangle &=& \epsilon \vert \psi_e^{-} \rangle \\
\hat{H}_{\text{qb}} \vert \psi_e^{-} \rangle &=& \epsilon \vert \psi_e^{+} \rangle.
\end{eqnarray}
If the measurement outcome is the even current, the qubits will remain in the even parity subspace during the entire measurement process but they will undergo oscillations between the even Bell states with a frequency $\epsilon/\hbar$, whatever the complex coefficients $\alpha, \beta, \gamma, \delta$ are. It is not possible to determine in which even Bell state the DQDs are entangled, they will be in a superposition state of $\vert \psi_e^+ \rangle$ and $\vert \psi_e^- \rangle$.\\

As an illustration of these properties, we consider two product states as initial states defined by $\alpha, \beta, \gamma, \delta = \pm 1/\sqrt{2}$ and we will compare their probabilities. These states are not only interesting theoretically \cite{Nielsen00} but they also present the strong advantage to be implementable with the experimental state-of-the-art of solid-state quantum dots.\cite{Hayashi03} The first state corresponds to $\alpha = \beta = \gamma = \delta = 1/\sqrt{2}$ whereas the second one corresponds to $\alpha = \beta = \gamma = -\delta = 1/\sqrt{2}$:
\begin{eqnarray} \label{DQDs_0}
\vert \Psi_1 \rangle &=& \frac{1}{2} \left( \vert {\uparrow}\rangle + \vert {\downarrow}\rangle \right) \otimes \left( \vert {\uparrow}\rangle + \vert {\downarrow}\rangle \right), \\
\vert \Psi_2 \rangle &=& \frac{1}{2} \left( \vert {\uparrow}\rangle + \vert {\downarrow}\rangle \right) \otimes \left( \vert {\uparrow}\rangle - \vert {\downarrow}\rangle \right)
\end{eqnarray}
It is straightforward to see that $P_e = P_o = 1/2$ in both cases. As expected the even probabilities exhibit an oscillation behaviour between the two even Bell states. The probabilities for $\vert \Psi_1 \rangle$ are:
\begin{eqnarray}
P_e^+(t) &=& \frac{1}{2} \cos^2\big(\frac{\epsilon t}{\hbar}\big),  \\
P_e^-(t) &=& \frac{1}{2} \sin^2\big(\frac{\epsilon t}{\hbar}\big),
\end{eqnarray}
and the probabilities for $\vert \Psi_2 \rangle$ are almost the same, they just differ by a $\pi/2$-phase.
In contrast the probabilities in the odd subspace are time independent and allow to distinguish between the two odd Bell states. Starting with $\vert \Psi_1 \rangle$, one finds:
\begin{equation}
P_o^+ = 1/2, \quad P_o^- = 0,
\end{equation}
whereas starting with $\vert \Psi_2 \rangle$, one gets:
\begin{equation}
P_o^+ = 0, \quad P_o^- = 1/2.
\end{equation}
This example clearly shows that the electronic MZI acting as parity meter entangles the qubits with probability one ($P_e + P_o =1$) and allows to know exactly in which odd Bell state the qubits are if the  current $I_o$ is measured. In a more general way, if one writes the initial coherent superposed state for both qubits as
\begin{equation}
\vert \Psi_{\text{in}} \rangle = \frac{1}{2} \left( \vert {\uparrow}\rangle + e^{i\eta_1}\vert {\downarrow}\rangle \right) \otimes \left( \vert {\uparrow}\rangle + e^{i\eta_2}\vert {\downarrow}\rangle \right), 
\end{equation}
where $\eta_1$ and $\eta_2$ are phases specific to the upper and lower qubit respectively, then the class of states that allows deterministic generation of odd Bell states is defined by:
\begin{equation}
\eta_2 -\eta_1 = 0 \quad \text{modulo} \quad \pi.
\end{equation}

This property makes it a very interesting quantum detector and entangler for further experiments.

\section{Conclusion}

With this work, we present an investigation of the MZI as quantum detector of two DQDs characterized by a bias and no tunneling between the states. Thanks to the control of the flux and of the transparency of the QPCs that can be tuned to specific values, the MZI operates as ideal detector, efficient parity meter and quantum entangler for the DQDs. Considering the qubits in an initial superposed state, the measurement process drives them into the Bell states. These states are of great interest for quantum information as they are the maximal entangled states for two qubits. This set of results encourages the implementation of such a setup in mesoscopic physics but also in other promising fields for quantum information such as circuit quantum electrodynamics. 

\section*{Acknowledgements}

We acknowledge a clarifying discussion with A.N. Jordan and we thank S.E. Nigg, Y. Blanter, L. DiCarlo and A. Baas for useful discussions. This work was supported by the Swiss NSF and MANEP.

\section{Appendix: Dephasing rates}

For calculating the dephasing rates given in Eqs.\ (\ref{rates1}-\ref{rates2}), two approaches are possible: the entanglement approach (Section A) or the charge fluctuations approach (Section B). The entanglement approach \cite{Clerk03, Hackenbroich00, Averin05} aims at writing the reduced density matrix for the qubits; the dephasing terms are extracted from the off-diagonal elements which are also called the coherence terms of the system. The charge fluctuations approach \cite{Buttiker00,Pilgram02,Clerk09} consists in investigating the effect of the transit of many electrons through the arms. This transport is stochastic and gives rise to fluctuating potentials in the arms also seen by the DQDs through the coupling. As a consequence the DQDs will exhibit a fluctuating phase which will lead to dephasing.

\subsection{Entanglement approach}

To get the information concerning the dephasing rates, we have to write the reduced density matrix of the two DQDs at time $t$, after the transit of N electrons. For this, we first consider the transit of one electron (one scattering event) in a time $\tau \ll t$, $\tau$ being the traversal time of one electron in the MZI. We assume that all consecutive scattering events are independent. As a consequence the entanglement generated by the measurement will be the product of the entanglement generated by each independent electron that transfers through the MZI during the  measurement time, $t = N\cdot \tau$. Initially the MZI and the two DQDs are completely disentangled:
\begin{eqnarray}
\vert \Psi_0 \rangle &=& \vert \Psi_{\text{det}} \otimes \vert \Psi^{qb}(0) \rangle \nonumber \\
&=& \vert \Psi_{\text{det}} \rangle \otimes \big(\alpha \vert{\uparrow}  \rangle + \beta \vert {\downarrow} \rangle\big)  \otimes \big(\gamma \vert{\uparrow}  \rangle + \delta \vert {\downarrow} \rangle\big),
\end{eqnarray}
where $\alpha, \beta, \gamma, \delta$ are the complex numbers defined in Section V and $\vert \Psi_{\text{det}} \rangle$ is the state of the detector before the experiment starts. When one electron transits through the arms of the MZI, it acquires information about the state of the DQDs. As a consequence the state of the detector and the state of the qubits become entangled. Assuming that the electron can go out in lead 3 and 4 of the MZI, the entangled state of the MZI and the qubits at time $\tau$ reads:\cite{Averin05}
{\small{\begin{eqnarray}
\vert \Psi(\tau) \rangle &=&\alpha \gamma \, e^{-i \epsilon \tau /\hbar}\,(S_{41}^{{\uparrow} {\uparrow}} \vert 4 \rangle + S_{31}^{{\uparrow} {\uparrow}} \vert 3 \rangle) \otimes \vert{\uparrow} {\uparrow} \rangle \nonumber \\
&+&\beta \delta \, e^{i \epsilon \tau /\hbar}\,(S_{41}^{{\downarrow} {\downarrow}} \vert 4 \rangle + S_{31}^{{\downarrow} {\downarrow}} \vert 3 \rangle) \otimes \vert{\downarrow} {\downarrow} \rangle \nonumber \\
 &+&\alpha \delta \,(S_{41}^{{\uparrow} {\downarrow}} \vert 4 \rangle + S_{31}^{{\uparrow} {\downarrow}} \vert 3 \rangle) \otimes \vert{\uparrow} {\downarrow} \rangle \nonumber \\
 &+&\beta \gamma \,(S_{41}^{{\downarrow} {\uparrow}} \vert 4 \rangle + S_{31}^{{\downarrow} {\uparrow}} \vert 3 \rangle) \otimes \vert{\downarrow} {\uparrow} \rangle.
\end{eqnarray}}}

$\vert 3 \rangle$ and $\vert 4 \rangle$ are the output detector states and $S_{ij}$ are the amplitudes of probabilities to go from input $j$ to output $i$. These amplitudes depend on the state of the DQDs through the phases $\chi_1$ and $\chi_2$ (see Eq.\ (\ref{S_31}) for the expression of $S_{31}$ for instance). The phase factor depending on the qubit bias $\epsilon$ describes the free evolution of the qubits during the time $\tau$. The reduced density matrix for the DQDs at time $\tau$ is given by
\begin{equation}
\rho^{qb}(\tau) = \langle 3 \vert \Psi(\tau) \rangle \langle \Psi(\tau)  \vert 3 \rangle +  \langle 4 \vert \Psi(\tau) \rangle \langle \Psi(\tau)  \vert 4 \rangle.
\end{equation}
The elements of this reduced density matrix can be expressed in the form:\cite{Hackenbroich00}
\begin{equation}
\rho^{qb}_{ij}(\tau) = \rho^{qb}_{ij}(0) e^{-i \omega_{ij} \tau} A_{ij}.
\end{equation}
For simplicity the density matrix is written in the computational basis $\{\vert {\uparrow} {\uparrow} \rangle, \vert {\uparrow} {\downarrow} \rangle, \vert {\downarrow} {\uparrow} \rangle, \vert {\downarrow} {\downarrow} \rangle \}$. $\rho^{qb}_{ij}(0)$ are the elements of the density matrix $\rho^{qb}(0) = \vert \Psi^{qb}(0)\rangle \langle \Psi^{qb}(0) \vert$. $A_{ij}$ depends on the state of the DQDs through $\Delta \chi_1$ and $\Delta \chi_2$ and on the reflection and transmission probabilities of the left QPC, $R_L$ and $T_L$. The frequency $\omega_{ij}$ depends on $\epsilon$ and describes free evolution corresponding to the matrix element $\rho_{ij}$. Then at time $t$, assuming that the N electrons are independent, the reduced density matrix's elements are given by:
\begin{equation}
\rho^{qb}_{ij}(t) = \rho^{qb}_{ij}(0) e^{-i\omega_{ij} t} A^N_{ij}.
\end{equation}
$t = N h / e V$ where e is the electrical charge and V is the applied potential between lead 1 and lead 3. That is to say $e V / h$ is the rate at which the electrons are sent into the interferometer or in an equivalent manner, it corresponds to the inverse of the traversal time $\tau$ introduced before. By comparing this expression for the off-diagonal elements with the standard expression of the coherence terms, $\rho^{qb}_{ij}(t) = \rho^{qb}_{ij}(0) e^{-i\omega_{ij} t} e^{-\Gamma_{\text{deph},ij}t}$, we deduce the expressions of the dephasing rates:
\begin{equation}
\Gamma_{\text{deph},ij} = - \frac{1}{t} \log\Big\vert A_{ij} \Big\vert^N.
\end{equation}

In the entanglement approach, two or more scattering matrices are used, depending on the number of qubits in the setup. In our case we deal with 4 scattering matrices namely depending on the charge state $\sigma \sigma'$ of the qubits, $S^{\sigma \sigma'}$. In contrast, in the charge fluctuations approach (see Appendix B), the scattering properties are expressed in terms of a reference scattering matrix $S$ (scattering matrix in the absence of the qubits) and in terms of functional derivatives with regards to potential variations. 
Thus in the entanglement approach, we can write:
\begin{equation}
S^{\sigma \sigma'} = S + \frac{\partial S}{\partial U} \delta U^{\sigma \sigma'} + \ldots,
\end{equation}
where $\delta U^{\sigma \sigma'}$ is calculated self-consistently \cite{Buttiker00, Pilgram02, ButtikerKu}. This provides the capacitance fraction in the entanglement approach. In our case the different scattering matrices differ only through phases, i.e $\Delta \chi = C_\mu / C_i$.  

As mentioned in Section IV, we obtain six different dephasing rates which reduce to the three given in Eqs.\ (\ref{rates1}-\ref{rates2}) when taking into account the parity assumptions, Eqs.\ (\ref{parity1}) and (\ref{parity2}).

\subsection{Charge fluctuations approach}
This approach investigates the fluctuating potentials in the arms of the MZI and in the DQDs, arising from the quantum statistical nature of electron transport. In the simple case of one two-level system coupled to a detector, the dephasing rate is given by \cite{Levinson97, Buttiker00, Pilgram02, Clerk09}
\begin{equation}
e^{-\Gamma_{\text{deph}}t} = \langle e^{i \delta \phi(t)} \rangle_U,
\end{equation}
where $U$ is the fluctuating potential of the detector and $\delta \phi(t)$ is the fluctuating phase acquired by the two-level system as a consequence of $U$. Assuming that $\delta \phi(t)$ is a Gaussian phase, the dephasing rate can be written as
\begin{equation}
e^{-\Gamma_{\text{deph}}t}  = \langle e^{i \delta \phi (t)} \rangle_U = e^{-\frac{1}{2}\langle \delta \phi(t)^2 \rangle_U},
\end{equation}
with $\langle \delta \phi(t)^2 \rangle_U \propto S_{UU}(0)\,t$ and $S_{UU}(0)$ is the spectral density of the voltage fluctuations at zero frequency. This spectral density can be expressed in terms of a non-equilibrium charge relaxation resistance \cite{Buttiker00,Pilgram02} $R_v$,  $S_{UU}(0) \propto \left( C_\mu/C_\text{i}\right)^2 R_v$. If only a single potential counts (spatial variations of the potential can be neglected),  the non-equilibrium charge relaxation resistance is given by: 
\begin{equation}
R_v = \frac{h}{e^2} \frac{\text{Tr}[N_{12}N_{12}^*]}{(\text{Tr}N)^2}, \quad N = \frac{1}{2 \pi i} S^\dagger \frac{dS}{dU}\frac{dU}{dE}.
\end{equation}
$N$ is a generalized Wigner-Smith matrix (a density-of-states matrix)  that is determined by the region inside the detector where the potential $U$ is non zero. In the case of one qubit coupled to a one-channel detector, \cite{Pilgram02} $N$ is a $2\times 2$ matrix and $N_{12}$ is its off-diagonal element.\\

Adapting this method to the case of two DQDs coupled to the MZI, we find the dephasing rates given in Eqs.\ (\ref{rates1}-\ref{rates2}). We assume two fluctuating potentials induced by the transfer of the electrons, $U_1(t)$ on the upper arm and $U_2(t)$ on the lower arm. Through the coupling capacitances $C_1$, both DQDs will acquire a fluctuating phase on top of their free evolution phase, $\delta \phi_1(t)$ for the upper DQD and $\delta \phi_2(t)$ for the lower DQD, which will depend on the state of the DQDs. Indeed there is no interaction when the charges in the DQDs are in the dot far from the arms. As in the simple case presented above, the correlators $\langle \delta \phi_i(t)\delta \phi_j(t) \rangle_U$ are proportional to the voltage fluctuations $S_{U_i U_j}(0)$ which are themselves proportional to the non-equilibrium charge relaxation resistance $R_v^{(ij)}$ given by:
\begin{equation}
\!R_v^{(ij)} \!=\! \frac{h}{e^2} \frac{\text{Tr}[N^{(i)}_{12}N^{(j)*}_{12}]}{\text{Tr}N^{(i)}\text{Tr}N^{(j)}}, \quad N^{(i)} \!=\! \frac{1}{2 \pi i} S^{(i)\dagger} \frac{dS^{(i)}}{d\chi_i}\frac{d\chi_i}{dE_i}.
\end{equation}
The phases $\chi_1$ and $\chi_2$ are those defined by Eqs.\ (\ref{phase1}) and (\ref{phase2}) and the Wigner-Smith matrix can be calculated from the scattering matrix given by Eq.\ (\ref{S_31}).

We find the following expressions for the correlators: 
\begin{eqnarray}
&& \langle \delta \phi_1(t)\delta \phi_2(t) \rangle_U = - R_L T_L \frac{eV t}{h} \left( \frac{C_\mu}{C_\text{i}}\right)^2, \\
&& \langle \delta \phi_1(t)^2 \rangle_U = \langle \delta \phi_2(t)^2 \rangle_U = R_L T_L \frac{eV t}{h} \left( \frac{C_\mu}{C_\text{i}}\right)^2.
\end{eqnarray}
These correlators show that the events on the upper arm and lower arm are perfectly anticorrelated as expected from the conservation of the current. Due to this perfect anticorrelation, one finds by calculating the six dephasing rates that they reduce to three:
\begin{eqnarray}
\!\Gamma_{eo}\!&=&\! \frac{1}{2t} \left\langle \delta \phi_{1,2}(t)^2 \right\rangle_U \!=\! \frac{1}{2} R_L T_L \frac{eV}{h}\! \left( \!\frac{C_\mu}{C_\text{i}}\!\right)^2,\\
\!\Gamma_{ee} \! &=&\! \frac{1}{2t} \left\langle \big(\delta \phi_1(t)\! -\! \delta \phi_2(t)\big)^2 \right\rangle_U \!=\! 2 R_L T_L \frac{eV}{h}\! \left(\! \frac{C_\mu}{C_\text{i}}\!\right)^2, \\
\!\Gamma_{oo}\! &=&\! \frac{1}{2t}  \left\langle \big(\delta \phi_1(t) \!+\! \delta \phi_2(t)\big)^2 \right\rangle_U \!=\! 0.
\end{eqnarray}
It is interesting to notice that the assumption $\Delta \chi_1 = \Delta \chi_2 = \Delta \chi$ is equivalent to assume a perfect anticorrelation between $\delta \phi_1(t)$ and $\delta \phi_2(t)$. These rates are exactly those obtained with the entanglement approach.

\end{document}